\input harvmac
\def\abstract#1{
\vskip .5in\vfil\centerline
{\bf Abstract}\penalty1000
{{\smallskip\ifx\answ\bigans\leftskip 1pc \rightskip 1pc 
\else\leftskip 1pc \rightskip 1pc\fi
\noindent \abstractfont  \baselineskip=12pt
{#1} \smallskip}}
\penalty-1000}
\def\fc#1#2{{#1\over #2}}
\def\frac#1#2{{#1\over #2}}

\def\br{\hfill\break}
\def\ni{\noindent}

%%%%%%%%%%%%%%%%%%%%%%%%%%%%%%%
\def\al{\alpha}
\def\eps{\epsilon}

\def\p{\partial}

\def\cx#1{{\cal #1}}
\def\tx#1{{\tilde{#1}}}

\def\us#1{\underline{#1}}
\def\hth/#1#2#3#4#5#6#7{{\tt hep-th/#1#2#3#4#5#6#7}}
\def\nup#1({Nucl.\ Phys.\ $\us {B#1}$\ (}
\def\plt#1({Phys.\ Lett.\ $\us  {B#1}$\ (}
\def\cmp#1({Comm.\ Math.\ Phys.\ $\us  {#1}$\ (}
\def\prp#1({Phys.\ Rep.\ $\us  {#1}$\ (}
\def\prl#1({Phys.\ Rev.\ Lett.\ $\us  {#1}$\ (}
\def\prv#1({Phys.\ Rev.\ $\us  {D#1}$\ (}
\def\mpl#1({Mod.\ Phys.\ Let.\ $\us  {A#1}$\ (}
\def\atmp#1({Adv.\ Theor.\ Math.\ Phys.\ $\us  {#1}$\ (}
\def\ijmp#1({Int.\ J.\ Mod.\ Phys.\ $\us{A#1}$\ (}
\def\jhep#1({JHEP\ $\us {#1}$\ (}

\def\subsubsec#1{\ \br \noindent {\it #1} \br}
%%%%%%%%%%%%%%%%%%%%%%%%%%%%%
%

%
\input epsf
\noblackbox
%
%\draftmode
%zdef%%%%%%%%%%%%%%%%%%%%%%%%%%%%%%%%%%%%%%%%%%%%

\def\vk3{V_{K3}}\def\msB{m_{susy}^{bulk}}\def\msb{m_{susy}^{brane}}

\def\mstr{M_{str}}\def\mpl{M_{pl}}\def\mew{m_{EW}}

\def\bb#1{{\bar{#1}}}

%enddefs%%%%%%%%%%%%%%%%%%%%%%%%%%%%%%%%%%%%%

\vskip-2cm
\Title{\vbox{
\rightline{\vbox{\baselineskip12pt\hbox{CERN-TH/2000-176}
\hbox{hep-th/0006204}}}}}
{Stringy World Branes and Exponential Hierarchies}
\abstractfont 
\vskip 0.2cm
\centerline{P. Mayr 
\foot{CERN Theory Division, CH-1211 Geneva 23, Switzerland}}

\abstract{%
We describe heterotic string and M-theory realizations of the 
Randall-Sundrum (RS) scenario with $\cx N=2$ and $\cx N=1$  supersymmetry in the bulk. 
Supersymmetry can be broken only on the world brane, a scenario that  
has been proposed to account for the smallness of the cosmological constant. 
An interesting prediction from string duality is the generation
of a warp factor for conventional type II Calabi--Yau 3-fold compactifications.
On the other hand we argue that an assumption that is 
needed in the RS explanation 
of the hierarchy is hard to satisfy in the string theory context.
}
\Date{\vbox{\hbox{ {June 2000}}
}}
\goodbreak
\parskip=4pt plus 15pt minus 1pt
\baselineskip=15pt plus 2pt minus 1pt
\leftskip=8pt \rightskip=10pt
\lref\lyk{J.D. Lykken, \prv 54 (1996) 3693.}
\lref\WitSc{E. Witten, \nup 471 (1996) 135.}
\lref\CHSW{P. Candelas, G.T. Horowitz, A. Strominger and E. Witten,
\nup 258 (1985) 46.}.
%\def\newsec#1{\ni{\bf #1}\br}
%%%%%%%%%%%%%%%%%%%%%%%%%%%%%%%%%%%%%%%%%%%%%%%%%%%%%%%%%%%%%%%
\newsec{Introduction}
An appealing interpretation of the hierarchy problem is in terms of a
world brane scenario with large extra dimensions that dilute gravity 
\WitSc\lyk%
\ref\AADD{
N. Arkani-Hamed, S. Dimopoulos and G. Dvali,
\plt 429 (1998) 263;
I. Antoniadis, N. Arkani-Hamed, S. Dimopoulos and G. Dvali,
\plt 436 (1998) 257.}.
It was further argued in 
\ref\RSi{L. Randall and R. Sundrum, \prl  83 (1999) 3370; \prl  83 (1999) 4690.},
referred to as RS in the following,
that the remaining problem of the unnatural
large extra dimension can be eased if the back-reaction of gravity
to the brane produces a total space-time with suitable
non-product structure.
Specifically the metric considered in these papers is of the warped form 
\eqn\warpm{
ds^2=e^{A(y)}\, \eta_{\mu\nu}dx^\mu dx^\nu+dy^2,
}
where $\mu,\nu=1,\dots,4$, $A(y)=-k|y|$ and $y$ takes values in a finite interval. 
Due to the exponential dependence of the 
warp factor on the transverse dimension, a moderate 
distance in the $y$ direction
\eqn\rsrs{\delta y \sim 2\, \ln  m/M}
may account for the large hierarchy $m_{EW}/\mpl$ 
between the scales of gravity and electro-weak interactions.

Note that there is an important assumption in this argument,
namely that the length scale $\delta y$ is the input which assumes
a natural value whereas the ratio of mass scales
$M/m$ is the derived quantity. In the framework of ref.\RSi, which is
supposed to be the effective description of a desired setup,
the validity of 
this assumption can not be decided.  If in contrary physics tells us 
that $M/m$ is the input quantity, naturally of order one, 
and $\delta y$ is a {\it derived} quantity, the logarithmic 
relation \rsrs\ is largely irrelevant for the hierarchy problem. 
We will argue that this is indeed the generic answer one finds in a string 
theory realization of this geometry. 
Nevertheless there are also indications that 
a sufficiently large interval $\delta y$ may be stabilized in the case of heterotic and
M-theory world branes.  

An universal warp factor present in any string theory is the
dilaton factor that relates the string sigma
model metric $g^{(s)}$ and the canonical Einstein metric
\eqn\stringm{
g_{MN}^{(E)}=e^{-2\alpha \phi}g^{(s)}_{MN},
}
where $\alpha$ is a positive number.
It follows that a compactification 
on a manifold $X$ with a dilaton that depends on the 
internal dimensions gives rise to a warp factor in space-time. If there are 
degrees of freedom localized on a ``world brane'', they
may probe this warp factor and the relation that links the  masses
as measured in the brane and bulk metric is $m_{brane}=e^{-\alpha \phi} m_{bulk}$.

In this note we argue that the world brane geometries of the RS type appear
naturally in heterotic string and M-theory vacua\foot{World brane 
embeddings in orientifold/F-theory have been discussed in
\ref\HVi{H. Verlinde, {\it Holography and compactification}, hep-th/9906182.}%
\ref\HVii{C.S. Chan, P.L. Paul and H. Verlinde, {\it
A note on warped string compactification}, hep-th/0003236.}%
\ref\GSS{B.R. Greene, K. Schalm and G. Shiu, {\it
Warped Compactifications in M and F Theory}, hep-th/0004103.}.
See also \ref\rBC{K.~Behrndt and M.~Cvetic, \plt 475 (2000)  253.}
for a related work in the effective supergravity.}
and study some of their
properties. Indeed the generic four-dimensional heterotic string vacua, 
the so-called (0,2) vacua, come with warp factors
\ref\AS{A. Strominger, \nup 274 (1986) 253.}%
\ref\WSD{B. de Wit, D.J. Smit and N.D. Hari Dass,
\nup 283 (1987) 165.}%
\ref\Witoi{E. Witten, \nup 268 (1986) 79.}. Moreover, on a subset
of the moduli space there are non-abelian gauge symmetries localized on a 
5-brane and the dilaton warp factor near the brane realizes the RS
geometry. 
So in some sense the condition for having a world brane of RS type 
is not (much)  more special than having non-abelian gauge symmetries,
which is granted by standard model physics!
A similar situation arises for 5-branes in M-theory, a subclass of which 
can in fact be considered as the strong coupling description of the heterotic 5-branes
via M-theory/heterotic duality 
\ref\HW{P, Horava and E. Witten, \nup 460 (1996) 506; \nup 475 (1996) 94.}.

The heterotic and M-theory brane worlds are interesting examples
of theories that provide equivalent dual descriptions in terms of either a gravity free
world volume theory or on the other hand a string theory in a geometric background 
\ref\mal{J. Maldacena, \atmp 2 (1998) 231.}%
\ref\GKP{S.S. Gubser, I.R. Klebanov and A.M. Polyakov, \plt 428 (1998) 105;\br
E. Witten, \atmp 2 (1998) 253.}%
\ref\SW{L. Susskind and E. Witten, {\it The holographic bound in anti-de Sitter space}, 
hep-th/9805114.}%
\ref\ABKS{O. Aharony, M. Berkooz, D. Kutasov and N. Seiberg, \jhep 10 (1998) 4.}.
The latter describes the higher-dimensional theory in which the brane is embedded
as a soliton. This dual perspectives have led to an intriguing proposal
\ref\HViii{H. Verlinde, {\it Supersymmetry at large distance scales},
hep-th/0004003.} that could explain the observed smallness of the cosmological 
constant
\ref\CS{C. Schmidhuber,  {\it 
Micrometer Gravitinos and the Cosmological Constant}, hep-th/0005248.}, 
if supersymmetry is broken only on the brane.
We find that the heterotic brane worlds
realize this situation in a very natural way.

\newsec{Heterotic $SO(32)$ small instantons}

\subsec{The warp factor}
Let us start with the small instanton in the heterotic SO(32) 
string \ref\EWsi{E. Witten, \nup 460 (1996) 541.},
which has a representation as a 5-brane soliton in the low energy supergravity theory 
\ref\DL{M.J. Duff and J.X. Lu, \nup 354 (1991) 141;
A. Strominger, \nup 343 (1990) 167;
C.G. Callan, J. Harvey and A. Strominger, \nup 359 (1991) 611.}\foot{This 5-brane can be
visualized as the 5+1 dimensional hypersurface transverse to a 4-plane $H$ with 
$\int_H F\wedge F=N$ and localized at the center of the instanton in $H$ in the zero size limit.}.
On its world volume lives an $\cx N=1$ supersymmetric $SU(2)$ gauge theory with sixteen matter multiplets in the 
fundamental representation. In the following we take this gauge theory on the brane as a candidate
to describe part of the standard model; more
general and realistic world volume theories will be mentioned later on.

If the 5-brane is embedded in flat ten-dimensional space-time, the six-dimensional
Planck mass on the brane is infinite. For this reasons we will consider the situation where 
the transverse dimensions are compactified on a 4-manifold $Y$. It is important however that 
the gravity on $Y$ is not of the standard form but highly altered by the presence of the 5-brane. 

Specifically is well-known \AS\WSD\Witoi\ that the generic heterotic 
string compactification
comes with a warp factor which is entirely due to a variation of the dilaton as in \stringm.
The warp factor does in general not lead to exotic effects for the low 
energy theory as long as the latter depends only on the quantities averaged over 
the internal manifold. A more interesting story arises in the presence of 
the small instantons. Firstly the small instantons 
introduce the localized degrees of 
freedom that probe the warp factor. 
Secondly they lead to an extreme local behavior of the supergravity fields and
especially the warp factor. 

In the presence of small instantons,  the Bianchi identity for the 
anti-symmetric tensor field strength becomes
\eqn\bianchi{
dH=-{  tr} R\wedge R+\fc{1}{30}{{  Tr} F\wedge F}+\sum_{i=1}^N \hat\delta_{5B},}
where $\hat\delta$ denotes a 4-form supported on 
the submanifold on which the $N$ instantons localize. 
The first two terms on the right hand side of eq.\bianchi\ are forced by anomaly
cancellation in the ten-dimensional heterotic string  
\ref\GS{M. Green and J. Schwarz, \plt 149 (1984) 117;\plt 151 (1985) 21.}.
The last term describes the small instanton. It displays
the important fact that the latter  contribute to the 
cancellation of gravitational anomalies localized on a 5-brane in space-time 
\ref\DMW{M.J. Duff, R. Minasian and E. Witten, \nup 465 (1996) 413}.
The integral over the right hand side must vanish for any 4-dimensional compact 
submanifold $Y$ in space-time:
\eqn\globconstr{
c_2(Y)-c_2(F)-N=0.}

The equation that determines the local variation of the dilaton - and thus the warp factor - 
near a generic point in the transverse directions, must be of the 
same form as that derived in \AS\ for a transverse K3 space:
\eqn\dileqii{
\Delta e^{2\phi}=-{ tr} R^2+\fc{1}{30}{  Tr} F^2+\sum_{i=1}^N \delta_{5B}.}
Locally the solution for the dilaton and the 
metric will be approximated by the neutral 5-brane solution of refs.\DL.
In the string frame one gets for $N$ small instantons located at a point
$r=0$:
\eqn\fbm{
ds^2=\eta_{\mu\nu}dx^\mu dx^\nu+e^{2\phi}\delta_{mn}dy^mdy^n,
\qquad e^{2\phi}=e^{2\phi_0}\, (1+\fc{N\al'}{r^2}),}
where $x^\mu,\, \mu=1,\dots,6$ and $y^m,\, m=1,\dots,4$ are coordinates in the 
dimensions parallel and transverse to the 5-brane, respectively.
To obtain a four-dimensional gauge theory on the world volume,
two tangential directions of the 5-brane
will be compactified on a complex curve $B$ with coordinate $z$. 
Introducing a new coordinate $y=\ln\, r$, the Einstein metric close to $r=0$ becomes:
\eqn\fbme{ds^2=\, c^{-1/2}e^{y}\eta_{\mu\nu}dx^\mu dx^\nu+c\, dy^2 \ +\ \{g_{z\bb z}(B)\ 
dz\, d\bar{z} \ 
+c\, d\Omega_3\},}
where $\mu,\nu=1,\dots,4$, $c=e^{2\phi_0}N$ 
and $d\Omega_n$ is the metric on the $n$-sphere.
Note that the first bracket describes a five-dimensional space with a metric of the
RS form \warpm\ with the world brane located at $y=-\infty$. 
The second bracket contains the metric for the additional transverse and compactified 
parallel dimensions which do not interfere with the warped five-dimensional 
geometry.

\subsubsec{Localization of gravity}
Let us compare the global structure of the warp factor as constrained by eq.\globconstr\ 
with that in the RS proposals \RSi. The first of these paper assumes localization of gravity on a 
hidden
``Planck brane'' with a sink in the warp factor at the world brane,
whereas the second describes localization of the graviton at a single brane. 
The behavior of the metric \fbme\ of the small instanton is of the first type.

Moreover the non-trivial structure of gravity in the transverse dimensions, namely
peaks and sinks for the warp factor corresponding to localization or
dilution of the graviton wave functions, is obviously characterized by the
non-vanishing of the term $c_2(Y)$. Since it is the integral of the ten-dimensional
counter term for gravitational anomalies, it depends on the transverse manifold Y.
Specifically, if $c_2(Y)=0$, then either there is a world-brane
and the space has to be non-compact to be consistent with \globconstr\ - which makes
gravity on the six-dimensional brane trivial.  Or $Y$ is a flat torus $T^n$ and
the six-dimensional Planck mass is finite, 
but the exponential sink at the world brane is prohibited by the 
the absence of appropriate compensating sources on the compact manifold. 

On the other hand, if $Y$ is a compact space with $c_2(Y)\neq 0$, 
the global condition \globconstr\ allows for the presence of 5-branes. The metric is locally determined by \fbme,
with an exponential sink at the world brane, and gives a perfect string theory embedding of the RS scenario. 
The Planck brane is replaced by nearby peaks in the graviton wave-function which are related to the
non-vanishing value of the anomaly counter term evaluated on $Y$. A qualitative picture is shown in 
Fig. 1.
\vskip0.25cm

{\baselineskip=12pt \sl
\goodbreak\midinsert
\centerline{\epsfxsize 4.35truein\epsfbox{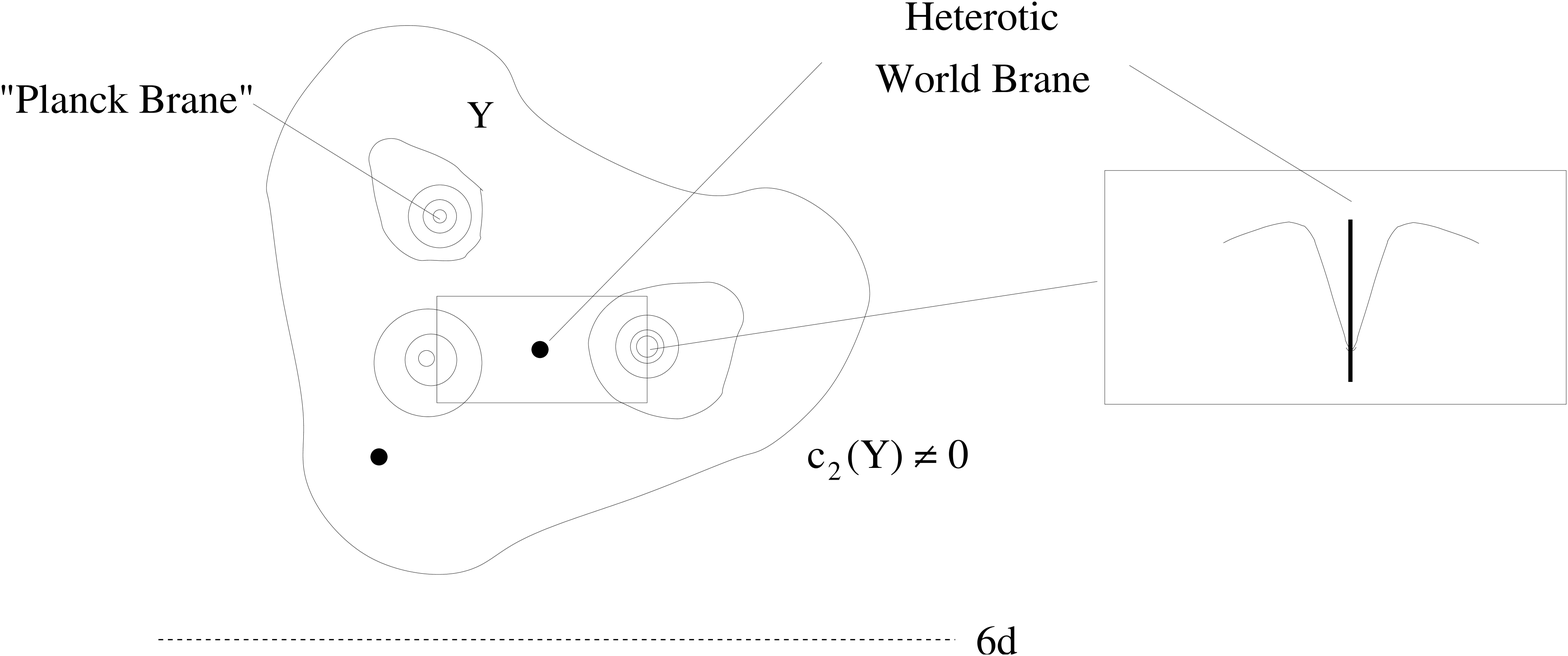}}
\leftskip 1pc\rightskip 1pc \vskip0.3cm
\noindent{\ninepoint  \baselineskip=8pt 
\vskip 0.4cm\ni
{{{\bf Fig. 1}: A sketch of the shape of gravity on the manifold $Y$. The local structure 
near the small instanton is identical to the RS model, while globally the Planck brane is 
replaced by sources of the graviton wave function arising from a non-zero value of 
the ``anomaly counter term'' $c_2(Y)$.}}}
\endinsert}

\subsubsec{Scales and couplings}
The simplest case of a transverse compact space $Y$ with $c_2\neq 0$
is a compact K3 manifold with $c_2=24$. 
To obtain a four-dimensional gauge theory 
the six-dimensional world volume will be further 
compactified on a complex curve $B$.
For the moment we assume that $B$ is a $T^2$ which gives the total manifold
a product structure $K3\times T^2$. The following discussion of scales 
and couplings generalizes straightforwardly to the
more general compactifications discussed in sect. 4.

The six-dimensional gravitational coupling constant is given by 
\eqn\sdpl{
\mpl^4=\mstr^8\ \int_Y \sqrt{g(Y)}e^{-2\phi} = e^{-2\phi_0}\, \mstr^4,\qquad
e^{-2\phi_0}=\fc{\mstr^4\, V_Y}{\lambda_H^2},}
where $\lambda_H$ is the ten-dimensional dilaton and $V_Y$ is the volume of K3 
without the warp factor in string units.  
In the second relation we have used the fact that we can calculate
the value of the integral for a configuration with constant dilaton. 
Moreover, in the string frame,  the value of the six-dimensional 
gauge couplings in the perturbative gauge sector and on the brane are \EWsi:
\eqn\sdgc{\fc{1}{g_{6,p}^2} \sim \mstr^2\, e^{-2\phi_0},\qquad
\fc{1}{g_{6,b}^2}\sim \mstr^2.}
Note that, contrary to what happens for the perturbative gauge groups where
one has the relation $g_6^2\mstr^2\sim (\mstr/\mpl)^4$, the string scale can be
arbitrarily small for a fixed gauge coupling and fixed Planck mass by
making the volume $V_Y$ large or the coupling $\lambda_H$ small.
This has led to the idea of weak scale superstrings \lyk\WitSc\foot{The
phenomenological viability of the scenario of \lyk\ has been 
stressed in 
\ref\rBO{K.~Benakli and Y.~Oz, \plt 472 (2000) 83.}.}.
They realize a world localized on a 
brane with gravity diluted by extra large dimensions, an idea which is also central to 
refs.\AADD. The four-dimensional couplings and scales follow from a 
trivial multiplication of the six-dimensional quantities with
the volume of $B$, as the world volume fills the entire six-dimensional space time.

However, the above discussion neglects the warp factor. 
Even if $\mstr\sim\mpl$, that is there is no large dimension, the hierarchy 
can be generated by an exponential warp factor 
$\eps\equiv e^{-\delta y/2}\sim m_{EW}/\mpl$ in a finite interval $\delta y$. 
From eq.\fbm\ it follows that 
the upper end of this interval, which we may choose to 
correspond to $y=0$, is set by a multiple of the string scale $\mstr$. 
The hierarchy $\eps$ is then determined by the lower end,
or IR cutoff, $y=-\delta y$. We will comment on the question to which extent a large 
interval $\delta y/2\sim 20$ is natural and stable in the present situation in sect. 6. 
Of course, more generally the hierarchy can also arise from 
a combined effect of the two mechanisms, with a moderately large dimension 
leading to an intermediate scale $\mstr< \mpl$ and the remaining hierarchy produced
by a warp factor.

\newsec{The cosmological constant and supersymmetry breaking}
Irrespectively of the hierarchy problem, there are quite interesting features
of the identification of the (heterotic) world brane theories with the observable world. 
An intriguing property is that a change in the vacuum energy of the world brane fields
does not change the effective four-dimensional cosmological constant, 
differently then what happens, at last naively, for the perturbative, non-localized
heterotic gauge sector.
Indeed the effective cosmological constant $\Lambda_4$
gets contribution from both the brane vacuum energy and the variation of the 
metric transverse to the brane and its total value is given by an integration constant $c$:
\eqn\ccs{
\Lambda_4=\lambda_{brane}+\lambda_{bulk}+(\p_y\phi)^2=c,}
where $\lambda_{brane}$ and $\lambda_{bulk}$ are the contributions from the
brane and bulk fields other than the metric, respectively.
For any chosen value of $c$, a change in the vacuum energy of the brane is canceled 
by the appropriate back-reaction of the transverse metric such that $\Lambda_4$ is kept fixed.
This is the mechanism of Rubakov and Shaposhnikov
\ref\RS{V.A. Rubakov and M.E. Shaposhnikov, \plt 125 (1993) 139.}
which was rediscovered in the context of 
AdS/CFT dualities in refs.
\ref\VVS{E. Verlinde and H. Verlinde, \jhep 05 (2000) 034;
C. Schmidhuber, {\it AdS(5) and the 4d cosmological constant}, hep-th/9912156.}.

As a consequence, the cosmological constant depends only on the initial value of $c$,
but not on
phase transitions or quantum corrections in the gauge theory on the brane. In 
the present context, a supersymmetric string
compactification to a maximal symmetric non-compact space-time $M$ requires $M$ to be 
Minkowskian \CHSW\AS\ and so $c=0$. In other words, requiring supersymmetry of the {\it bulk}
theory, selects the appropriate boundary condition such that $\Lambda_4$ is zero, irrespectively
of the vacuum energy of the world brane fields.

The interesting question however is what determines the size of $\Lambda_4$ 
when supersymmetry is broken. If supersymmetry is, in first approximation, only
broken on the brane, then the above mechanism still works to cancel the
effective cosmological constant from a brane vacuum energy
by an adjustment of the transverse metric. A non-zero 
cosmological constant arises therefore only from the sub-leading effect that the
cancellation mechanism will be disturbed by 
the propagation of supersymmetry breaking into the bulk. 
Thus the leading order contribution to $\Lambda_4$ is determined by the scale $\msB$, 
rather then $\msb$, which characterizes the mass splittings
in the observable sector. 

Now morally the hierarchy between the  mass scales of the bulk and 
the brane arises from the fact that the interactions between the two 
type of fields are suppressed by the warp factor $\eps=\mew/\mpl\sim \msb/\mpl<<1$. 
The idea of ref.\HViii\ 
is that if a similar factor suppresses the propagation of supersymmetry breaking back 
into the bulk, this may lead to a sufficiently small value of $\Lambda_4$. 
In fact it has been argued in ref.\CS\ that the 1-loop contribution to the cosmological
constant induced by the supersymmetry breaking in the bulk leads to values close to 
the observed data.

To introduce a mechanism that breaks supersymmetry exclusively on the brane is not trivial,
however. In the holographic correspondence the mass operator on the brane is dual to 
the vev of a supergravity bulk field. Generically, an operator that leads to a 
mass $m$ on the brane corresponds  to a mass scale $M$ in the bulk that is {\it larger},
$m^2=e^y\, M^2$. It is therefore unclear whether it is even in principle possible to 
obtain an appropriate warped geometry that describes a supersymmetry breaking 
sufficiently confined to a neighborhood of the brane in the pure supergravity sector.

An extra sector that comes for free in the case of the heterotic world branes
and seems to be tailor-made to realize the above situation 
is the perturbative gauge sector of the heterotic 
string with gauge group $G_0=SO(32)$.
In  an $\cx N=1$ supersymmetric vacuum,  strong gauge interactions 
in this string sector lead to a fermion condensate 
at a scale $\Lambda_F\sim e^{-8\pi^2/bg^2_{4,p}}$ 
that may break supersymmetry non-perturbatively
\ref\GC{For a review and original references, see
H.P. Nilles, \ijmp 5 (1990) 4199.}.
Let us assume that this scale is very small in the bulk, as a consequence of a 
small asymptotic value of the dilaton away from the world branes or since $G_0$
is broken to a subgroup with small $b$, or a combination of these effects.
No matter how small the mass scale $\Lambda_F$ is away from the 
five-branes it will grow with decreasing distance and induce
a large scale $\Lambda_F$ of the order of the string scale 
in the $G_0$ gauge theory near the brane
\eqn\gcond{\Lambda_F(y)\, \sim \, (\Lambda_F)^{exp(-2\phi)}\ \ \ \
%^{{y\to -\delta y}\atop\longrightarrow}
_{\longrightarrow} \! \! \! \! \! \! \! \! \! \! \! \! \! ^{y\to -\delta y}\ 
\ (\Lambda_F)^{{\eps}^4}\, \sim \, 1,
}
in units of $\mstr$.
Supersymmetry breaking is therefore restricted to the immediate local 
neighborhood of the world branes.

Although the process of supersymmetry breaking is non-perturbative in the heterotic 
theory and therefore difficult to access, one can, in favorable cases, 
use heterotic/type II duality to determine these effects by mirror symmetry 
\ref\PMssb{P. Mayr, {\it On supersymmetry breaking in string theory and its
                  realization in brane  worlds}, hep-th/0003198.}. We will discuss 
the heterotic/type II duality in presence of the world branes 
in more detail in the next section.%

\ni\vskip 0.5cm
\vbox{%\midinsert
\centerline{\epsfxsize 2.4truein\epsfbox{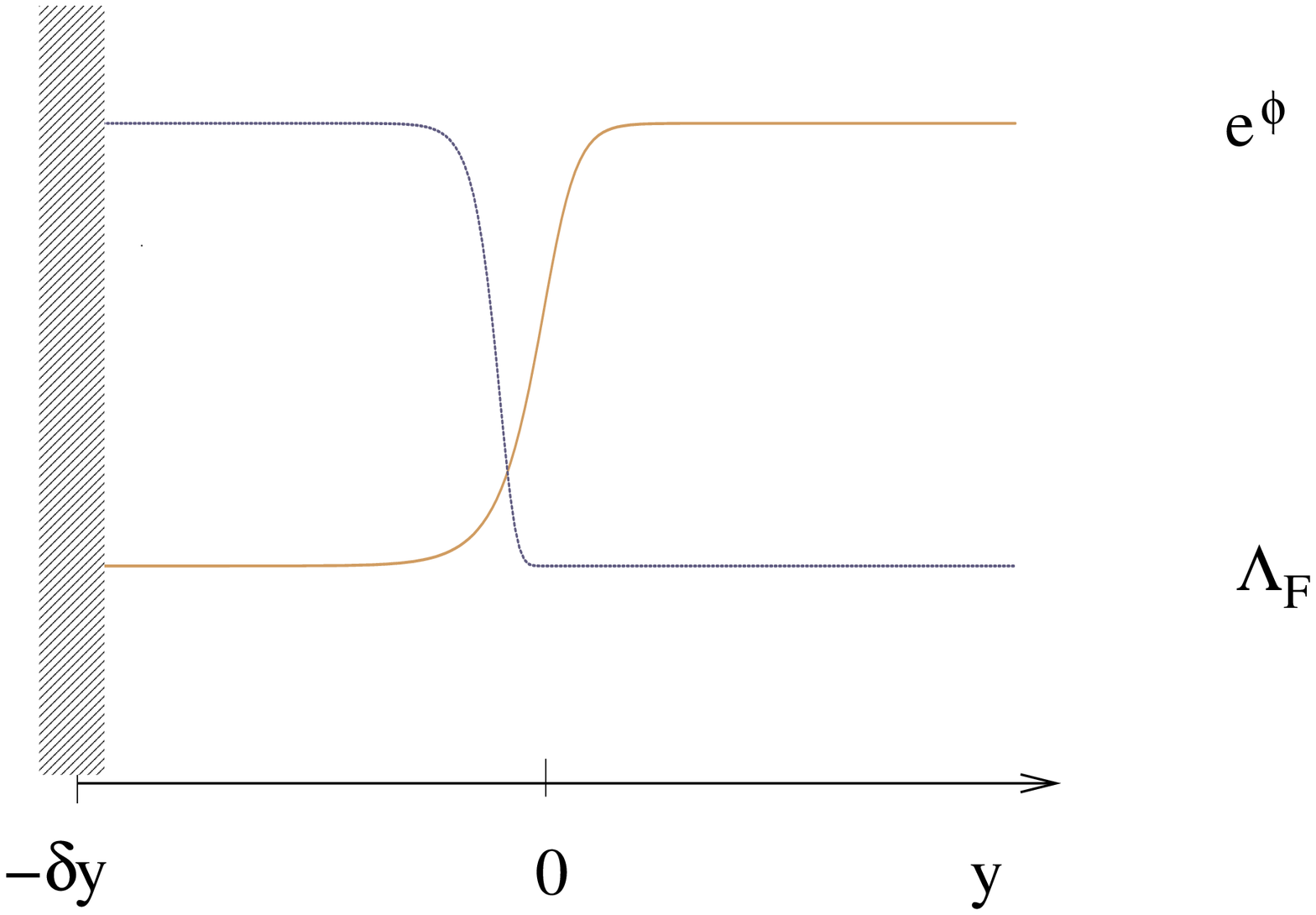}}
\leftskip 1pc\rightskip 1pc \vskip0.3cm
\noindent{\ninepoint  \baselineskip=8pt 
\vskip 0.1cm\ni
{{{\bf Fig. 2}: The behavior of the warp factor $e^\phi$ 
and the mass scale $\Lambda_F(y)$ of the fermion condensate
in the perturbative gauge sector 
as a function of the distance from the brane $y$. The warp factor decays 
exponentially towards the brane located to the left. In contrary the
gaugino condensate increases to the value $\Lambda_F(-\delta y) 
\sim \mstr$ in the neighborhood of the brane.}}}
%\endinsert
}

\newsec{Brane compactifications and string duality}
We proceed with a short description of four-dimensional compactifications with 5-branes 
with $\cx N=2$ and $\cx N=1$ supersymmetry and discuss some
of their properties.  An interesting implication arises from
heterotic/type II duality.

\subsec{Four-dimensional compactifications with $\cx N=2$ and $\cx N=1$ supersymmetry}
For simplicity we will mention in the following mainly the $SU(2)$ world volume theory of 
one small instanton at a generic point in the transverse space-time.
Obtaining a phenomenological viable spectrum 
is, as usual, a question of model building. However it is important to note
that the available spectra are sufficiently general to embed the
standard model in many ways.
Specifically more general world volume theories arise for $N$ coinciding 
instantons and instantons at non-generic singular points in the
transverse space-time. A brief summary of, and references on, 
these spectra are included in the appendix.

\subsubsec{Brane worlds with $\cx N=2$ supersymmetry}
The simplest case of a four-dimensional compactification with 5-branes is the manifold $X$ 
conformal to K3 times an extra $T^2$. In addition to the $SU(2)$ world volume gauge theory
with matter there is a single hypermultiplet for each 5-brane that parametrizes its
position on K3. Moreover, in four-dimensions the gauge theory is $\cx N=2$ supersymmetric
with the extra scalars in the vector multiplets representing the Wilson lines on 
$T^2$. There are also Wilson lines in the perturbative $SO(32)$ gauge factor.
They parametrize the masses of the fundamental hypermultiplets in the world volume
theory.

The compactification on $X$ with non-trivial dilaton in the absence of
5-branes has been studied in \AS\ and it is entirely straightforward
to adapt the discussion to include the 5-branes. 
As $c_2(X)=24$, the global constraint \globconstr\ can be  
satisfied by taking a trivial gauge background $F$ and adding
24 small instantons transverse to K3. The precise warp factor is determined by \dileqii\ with
the Laplacian understood to act only in the four directions of K3. 
An explicit expression for the curvature terms can be given in the orbifold limit.
Including the small instantons, the equation for the warp factor becomes
\eqn\ofl{
\Delta \fc{e^{2\phi}}{8\pi^2}=-\fc{3}{2}\sum_{k=1}^{16}\delta^4(x-x_k)+
\sum_{i=1}^{24} \delta(x-\tx x_i).}
Here $x_k$ are the positions of the 16 orbifold fixed points at which
the curvature of K3 has been localized. Moreover the 
$\tx x_i$ are the locations of the the 24 5-branes.

\subsubsec{Brane worlds with perturbative $\cx N=1$ supersymmetry}
Four-dimensional theories with $\cx N=1$ supersymmetry are obtained
by compactifying the heterotic string on a Calabi--Yau 3-fold $X_3$
\CHSW.
Similar as before, evaluating $c_2(S)$ on 
a holomorphic four-dimensional submanifold $S$ in $X_3$ determines the possible
number of sinks due to world branes transverse to $S$. The cycle $C$
dual to $S$ is holomorphic, and thus a supersymmetric cycle,
on which the 5-brane of the small instanton may be wrapped
\WitSc. For a generic
position of the 5-brane the behavior of the local - or low energy - 
description of the warp factor is still described by \fbm.

The spectrum of the world volume theory of a 
small instanton wrapped on a general curve $C$ is
determined by the normal bundle of the curve\foot{See 
refs.
\ref\FBs{P. Berglund and P. Mayr, \atmp 2 (1999) 1207;
G. Rajesh, \jhep 12 (1998) 18;
D.-E. Diaconescu and G. Rajesh, \jhep 6 (1999) 2;
R. Donagi, B.A. Ovrut and D. Waldram, \jhep 11 (1999) 30;
P. Berglund and P. Mayr, \jhep 12 (1999) 9;
A. Grassi, Z. Guralnik and B. Ovrut, {\it Five-brane BPS states in heterotic M-theory},
hep-th/0005121.}
for a discussion of various aspects of such 5-brane wrappings.}. In 
particular the number of supersymmetries will depend
on the dimension of the moduli space of $C$; an isolated (non movable) 
cycle $C$ is related to $\cx N=1$ supersymmetry while 
wrapping on curves of genus $g=1$ may lead to $\cx N=2$ 
supersymmetric theories 
\ref\klmvw{A. Klemm, W. Lerche, P. Mayr, C. Vafa and N. Warner, \nup 477 (1996) 746.}.

\subsec{A puzzle from Heterotic type/IIA duality}
String dualities map the gauge sector of one theory to a gauge sector of in general 
completely different origin in another theory. It is far from being obvious that 
they will map a gauge theory on a world brane with exponential warp factor
to a localized gauge theory with a similar warp factor in the dual theory. 

E.g., the heterotic string has on one hand the perturbative, non-localized 
gauge theory with gauge group $G_0$
and on the other hand the gauge theories on the world branes with a warp factor discussed in this paper.
For a $K3\times T^2$ compactification this string is dual to a type IIA compactification
on a Calabi--Yau 3-fold $X$ \ref\KVp{S. Kachru and C. Vafa, \nup {450} (1995) 69;
S. Ferrara, J. A. Harvey, A. Strominger
                  and C. Vafa, \plt {361} (1995) 59.},
where {\it both} gauge symmetries arise from localized 
singularities \ref\EWii{E. Witten, {\it Some comments on string dynamics},
hep-th/9507121; M. Bershadsky, C. Vafa and  V. Sadov, \nup 463 (1996) 398;
P. Mayr and A. Klemm, \nup 469 (1996) 37;
S. Katz, D.R. Morrison and  M.R. Plesser, \nup 477 (1996) 105.}
and thus live on a brane.
No warp factor has been found so far in this type IIA
compactification. How can two such theories be dual ? The answer to this question is 
bound to be interesting:

\ni
{\it Resolution 1: No warp factor}: The first possibility is that the localized (heterotic)
world brane theory with a warp factor maps to a theory without a warp factor 
in the dual (type IIA) theory. In fact it has been
speculated for a while what might be the proper generalization of
the AdS/CFT duality to describe a four-dimensional world {\it
with} dynamical gravity, similar to ours. A proper description
of the field theory side coupled to gravity must involve string theory, so one expects
that AdS/CFT duality is replaced by a suitable string/string duality. In an
appropriate scaling limit one must recover string theory on 'AdS' on
the one side and the 'CFT' on the other side. If the type IIA
string metric remains its product form including $\al'$ and  quantum corrections,
it is likely to represent a string duality embedding of a 'AdS/CFT' correspondence.
In the appropriate limit, the type IIA string represents the decoupled
four-dimensional 'CFT' while the heterotic side describes its 
'AdS' dual.

\ni
{\it Resolution 2: A ``quantum'' warp factor}: The second possibility is 
that the localized theories with warp factor match 1-1 to each other under
the duality. In fact we want to argue that this is what happens in the
present case of the heterotic/type IIA duality. 

To see which effect in the type IIA theory might result in a so far neglected warp factor,
note that the terms on the right hand side of eqs.(2.1) and (2.3) 
are higher order effects in $\al'$
in the heterotic string. In fact precisely the same terms have played a crucial role in the
equivalence of the hypermultiplet moduli spaces of the heterotic string on K3 singularities of type $G$
and three-dimensional $G$ gauge theories with $\cx N=4$
supersymmetry \ref\Witade{E. Witten, {\it Heterotic string conformal field theory and A-D-E
singularities}, hep-th/9909229.}. In particular the $\al'$ correction to the heterotic
moduli maps to a 1-loop term in the 3d gauge theory. The isomorphicity of these moduli
spaces for general $G$ has been established in 
\ref\adesings{P. Mayr, {\it Conformal field theories on K3 and three-dimensional gauge
                  theories}, hep-th/9910268.} by showing that the gauge theory
comes to life in a type II compactification\foot{See also ref. \ref\MR{M. Rozali, \jhep 12 (1999) 13.}
for a M-theory derivation of the equivalence of these moduli spaces.} which is the string dual of 
the heterotic string on the K3 singularity times a 
$T^3$.

This suggests that a similar effect as the one that corrects the hypermultiplet moduli
of the type IIA theory in the above equivalence may also create a warp factor.
An important role must be played by the $R^4$ terms in the type II action 
which contribute
to the hypermultiplet metric and the Einstein term in four dimensions 
\ref\AFGM{I. Antoniadis, S. Ferrara, R. Minasian and K.S. Narain, \nup 507 (1997) 571.}.
Note that to reproduce the structure of warp factors in the heterotic string, 
the relevant corrections in the type II compactification should arise only from those
singularities that describe the non-perturbative world volume theories, but not from the others that 
support the gauge theories dual to the perturbative gauge group of the heterotic string.

An interesting consequence of such a  warp factor for the type IIA gauge 
theory would be that we could extend the geometric engineering approach of refs.\klmvw%
\ref\gerefs{S. Katz, A. Klemm and C. Vafa, \nup 497 (1997) 173; S. Katz, P. Mayr and C. Vafa, 
\atmp 1 (1997) 53.}
to describe AdS/CFT duality. In lowest order, 
the type IIA vacuum is of the form $X_3\times M_4$ and describes 
the (possibly conformal) 
field theory on flat $M_4$. Adding the warp factor should deform this geometry 
locally to one of the type $AdS\times Y_5$, where $Y_5$ is a 5-dimensional space. Note that there 
are no RR backgrounds involved and therefore the determination of $\al'$ corrections, which is
essential for field theories with a low amount of supersymmetry may be available by conventional
methods, such as mirror symmetry. 
More details will appear elsewhere 
\ref\wip{Work in progress.}.

\newsec{Small $E_8$ instantons and the M-theory 5-brane}
Instead of the small $SO(32)$ instanton, we can also consider the small $E_8$ instanton which is 
described by the same equations (2.1-2.5). The main difference is the spectrum on the world volume.
For a six-dimensional K3 compactification, the $SU(2)$ gauge theory is replaced by a  $\cx N=1$ tensor 
multiplet together with the hypermultiplet that describes the location on K3. The $B_{\mu\nu}$
field in the tensor multiplet has an anti-self-dual field strength and couples to  a tensionless string 
\ref\eefb{O.J. Ganor and A. Hanany, \nup 474 (1996) 122;
N. Seiberg and E. Witten, \nup 471 (1996) 121.}. A 
nice geometric description is in terms of the M-theory dual \HW\
on $S^1/Z_2\times K3$, where the 
small instanton corresponds to a 5-brane stuck on one of the two end of the world 9-branes. 
The tensionless strings on the 5-brane represent the boundary of a membrane ending on it.
Upon a compactification on a circle we obtain conventional gauge theories.
From the tensionless strings one obtains also matter multiplets that are charged under
the perturbative gauge symmetry of the heterotic string  \ref\BM{See first reference in \FBs.}.
In fact a large class of these theories will be equivalent to the theories from the small
$SO(32)$ instanton by the usual T-duality 
\ref\PG{P. Ginsparg, \prv 35 (1987) 648.}.

An interesting new branch in the moduli space of these theories 
is the one where the 5-brane leaves the 9-brane, which corresponds to 
a vev for the scalar in the tensor multiplet (and a Coulomb branch in the T-dual $SO(32)$ theory).
For $N$ coinciding 5-branes in the bulk compactified further on a circle one obtains a $SU(N)$
gauge theory on the world volume theory with one adjoint hypermultiplet 
\ref\ASii{A. Strominger, \plt 383 (1996) 44.}. Except for
the interactions with the 9-branes, which are suppressed by the distance to them, this 5-brane is the
same as the M-theory 5-brane. 

Similar as the small heterotic instanton \bianchi, the M-theory 5-brane contributes to gravitational anomalies. It also contributes 
to the Bianchi identity of the 4-form field strength $dF=\sum_k\hat{\delta}_k$.
The details of the anomaly cancellations are different 
\ref\EWmii{E. Witten, \nup 463 (1996) 383.}%
\ref\DLM{M.J. Duff, J.T. Liu and R. Minasian, \nup 452 (1995) 261.}
as the eleven-dimensional supergravity has no anomalies on a smooth manifold. For this reason the
5-branes arise in M-theory compactification with singularities, such as 
the $S^1/Z_2$ compactification 
and the orbifolds without a heterotic dual described in 
\EWmii%
\ref\mtofs{K. Dasgupta and S. Mukhi, \nup 465 (1996) 399; 
A. Sen, \prv 53 (1996) 6725.}.
These singularities contribute the necessary terms in the gravitational equations of motions
which take over the role of the term $c_2(X)$ in the case of the perturbative heterotic string.
The solution for the warp factor in a neighborhood of $N$  M-theory 5-branes is 
\ref\guv{R. Gueven, \plt 276 (1992) 49.}
\eqn\metiii{
ds^2=\triangle^{-1/3}\, \eta_{\mu\nu}dx^\mu dx^\nu+
\triangle^{2/3}\, \delta_{mn}dy^mdy^n,
\qquad \triangle=1+\fc{N\, \pi \, l_{11}^3}{r^3}.}
After compactification on the curve $B$ this is again of the RS form near the brane location $r=0$
\eqn\metiv{
ds^2=c^{-1/3}\, e^y\, \eta_{\mu\nu}dx^\mu dx^\nu+c^{2/3}\, dy^2+
\ (g_{z\bb z}(B)\ dz\, d\bar{z} \ +c^{2/3}\, d\Omega_4),}
with $c=\pi N$ and $\mu,\nu=1,\dots,4$.
The duality between the world volume theory and the supergravity or string background for this
case has been discussed in detail in \mal\ABKS.

At this point the reader might wonder 
why the exponential warp factor did not appear in the five-dimensional 
supergravity description of refs.\WitSc%
\ref\LOSW{A. Lukas, B.A. Ovrut, K.S. Stelle  and D. Waldram, \prv 59 (1999) 086001.},
which give a {\it linear} warp factor in the direction separating the 9-branes, usually denoted as $x^{11}$.
However, neither for the heterotic 5-brane which is stuck on the 9-brane, nor the M-theory
5-brane in the bulk, there is a reason to single out the $x^{11}$ direction near the 5-brane.
Locally the only natural direction is the radial distance $r$ from the brane in which 
the warp factor is of the exponential form described in eqs.\fbme\ and \metiv.

\newsec{Comments}

In this note we argued that world brane geometries  
of the Randall-Sundrum type appear naturally in 
generic heterotic string and M-theory compactifications with enhanced non-abelian gauge symmetries.
The heterotic and M-theory 5-branes give an interesting and natural example of
world branes with a holographic duality and dynamical gravity on the brane. The four-dimensional
theories obtained by compactifying two of the brane dimensions are interesting candidates for a 'standard
model' on the brane. In particular it will
be interesting to determine the structure of supersymmetry breaking for 
$\cx N=1$ compactifications and to compare the resulting value of the cosmological constant 
with the proposals in refs.\HViii\CS.

On the other hand, although the RS geometry appears naturally in string theory,
it does not necessarily lead to an explanation of the hierarchy: 
why should we consider the length scale 
$\delta y$ as more fundamental than the ratio of mass scales $m/M$? Note that a  
simple line of argument which would identify the distance of the branes with a 
modulus $\phi$ and then assign some natural vev to it applies equally well to the 
alternative identification of $\phi$ with the mass and so does not resolve the issue.

More concretely, the UV end, say at $y=0$,  of the interval $\delta y$ will be set by 
some multiple of the string scale $c\, \mstr$. Smaller
values of $y$ correspond to the IR flow of the world volume theory. If the world volume
theory is conformal, there is an infinitely long homogeneous flow to $y=-\infty$. 
On the other hand if the theory has a mass scale $m$, as it should if it serves
as a model for a world brane,
the exponential flow of the warp factor 
will stop at $y=2\, \ln m/\mstr$.
E.g. this is precisely 
what happens in the case of D3-branes,
which have an $\cx N=4$ supersymmetric 
world volume theory 
\ref\PS{
J. Polchinski and M.J. Strassler, {\it 
The string dual of a confining four-dimensional gauge
                  theory}, hep-th/0003136;
K. Pilch and N.P. Warner, {\it N = 1 supersymmetric renormalization group flows from IIB
                  supergravity}, hep-th/0006066.}. 
Operators in the world volume theory that break conformal invariance
and part of the supersymmetry, such as fermion masses, correspond linearly 
to vev's  of anti-symmetric tensors in the dual supergravity description.

Now the point is that the scale characterizing this anti-symmetric tensor backgrounds, 
as well as any other generic mass scale $m$ in the 
string theory, is naturally of the order of $\mstr$. Thus the length $\delta y$
of the exponential flow of the warp factor is a {\it derived} quantity, 
proportional to $\ln m/\mstr$.
In particular the natural value of the length scale $\delta y$ is 
very small in string units and does not 
account for the required hierarchy.  

It should therefore be clear that a situation in string theory, 
where $\delta y$ is naturally of order one must be a very peculiar one.
In this respect it does not seem that the exponential warp factor leads to a
substantial facilitation of the hierarchy problem in the well-defined context of
string theory.

However it is worth pointing out that the situation for the 
heterotic world branes is at least more promising than for D-branes
in this respect. The reason is that the mere existence of the world volume
degrees of freedom implies that however small the asymptotic string coupling
$e^{\phi_0}$ is away from the brane, the string coupling $e^\phi$ has to be at least of order one 
at the brane \EWsi. So the quantity 
$e^{\delta\phi}=e^{\phi-\phi_0}\sim \eps^{-2}$ which characterizes at the same
time the hierarchy between the scales on the brane and in the bulk, can be large\foot{A similar
conclusion has been reached in \PS\ for type II NS 5-brane solutions which are
related by $S$-duality to the perturbed D3 brane configurations.}. It would be interesting to study further 
whether this property of the world branes considered in this note 
may be used to make the RS argument work in a string theory context.

\vskip 1cm

\ni
{\bf Acknowledgments}: 
I would like to thank the CIT-USC center for theoretical 
physics for hospitality during the course of this work
and Wolfgang Lerche, Christof Schmidhuber, 
Stephan Stieberger and Nick Warner for valuable discussions. I am also grateful to 
Karim Benakli, 
Andreas Brandhuber, Renata Kallosh, Djordje Minic, Hans-Peter Nilles
and Yaron Oz for helpful conversations.

\appendix{A}{More references on more world volume theories}
For $N$ coinciding small $SO(32)$ instantons the world volume theory is 
a $Sp(N)$ $\cx N=1$ supersymmetric gauge theory in six dimensions 
together with a hypermultiplets in the $(32,2k)$ representation,
and an antisymmetric representation of $Sp(k)$ \EWsi. For $N$ coinciding 
small $E_8$ instantons one obtains a world volume theory of $N$ $\cx N=2$
tensor multiplets and in addition tensionless strings \eefb. Upon compactification
on a circle the tensor multiplets become vector multiplets and the strings
wrapped on $S^1$ particles. For a certain subset of the moduli space, the 
theory will be T-dual to the compactification of the $SO(32)$ small instantons.

A large variation of spectra arises if the $N$ instantons are moved to 
singular points in the transverse K3
\ref\SIOF{J.D. Blum
and K. Intriligator, \nup 506 (1997) 199; \nup 506 (1997) 223;
P.S. Aspinwall, \nup 496 (1997) 149; P.S. Aspinwall and D.R. Morrison,
\nup 503 (1997) 533;
P. Berglund and P. Mayr, \atmp 2 (1999) 1207;
P.S. Aspinwall, \jhep 4 (1998) 19;
P.S. Aspinwall and R.Y. Donagi, \atmp 2 (1998) 1041.}. 
In particular the $E_8$ instanton may lead to extra gauge degrees of freedom 
located at the singularity, while the $SO(32)$ instantons have now also
tensor multiplets in their spectrum. 

Upon compactification on a complex curve $B$, the spectrum on the world volume depends 
on the embedding of $B$ in the space-time and in particular on the number of 1-cycles $\gamma_i$ in
$H_1(B)$; reduction of a six-dimensional vector (tensor) on $\gamma_i$ leads generically to a scalar (vector).
The theory on the world volume is twisted 
\ref\BSV{M. Bershadsky, V. Sadov and  C. Vafa, \nup 463 (1996) 420.}
and the precise spectrum depends on the normal bundle to $B$ in space-time. See e.g.
\klmvw%
\ref\KSS{S. Kachru, N. Seiberg and E. Silverstein,
\nup 480 (1996) 170.}
for calculations of this type in special cases.

\listrefs

\end